\documentclass[aps,twocolumn,
superscriptaddress,
footinbib,
prl]{revtex4-2}

\usepackage{amsmath,amssymb,bm,bbold}
\usepackage{mathtools}
\usepackage{graphicx}
\usepackage{multirow}
\usepackage{epstopdf}
\usepackage{latexsym}
\usepackage[normalem]{ulem} 
\usepackage[caption=false]{subfig}
\usepackage[usenames,dvipsnames]{color}
\usepackage{hyperref}
\usepackage{xcolor}
\usepackage{braket}
\usepackage{physics}
\usepackage{stackengine}
\usepackage[utf8]{inputenc} 
\usepackage{natbib}

\begin{document}

\title{Floquet Hamiltonian Engineering of an Isolated Many-Body Spin System}

\date{\today}

\author{S.~Geier*}
\affiliation{Physikalisches Institut, Universit\"at Heidelberg, Im Neuenheimer Feld 226, 69120 Heidelberg, Germany}
\author{N.~Thaicharoen*$^{,\dagger}$}
\affiliation{Physikalisches Institut, Universit\"at Heidelberg, Im Neuenheimer Feld 226, 69120 Heidelberg, Germany}
\affiliation{Research Center for Quantum Technology, Faculty of Science, Chiang Mai University 239 Huay Kaew Road, Muang, Chiang Mai, 50200, Thailand}
\author{C.~Hainaut*}
\author{T.~Franz}
\author{A.~Salzinger}
\author{A.~Tebben}
\author{D.~Grimshandl}
\author{G.~Z\"urn}
\author{M.~Weidemüller$^{\dagger}$}
\affiliation{Physikalisches Institut, Universit\"at Heidelberg, Im Neuenheimer Feld 226, 69120 Heidelberg, Germany}

\begin{abstract}
Controlling interactions is the key element for quantum engineering of many-body systems. Using time-periodic driving, a naturally given many-body Hamiltonian of a closed quantum system can be transformed into an effective target Hamiltonian exhibiting vastly different dynamics. We demonstrate such Floquet engineering with a system of spins represented by Rydberg states in an ultracold atomic gas. Applying a sequence of spin manipulations, we change the symmetry properties of the effective Heisenberg XYZ Hamiltonian. As a consequence, the relaxation behavior of the total spin is drastically modified. The observed dynamics can be qualitatively captured by a semi-classical simulation. Synthesising a wide range of Hamiltonians opens vast opportunities for implementing quantum simulation of non-equilibrium dynamics in a single experimental setting.
\end{abstract}

\maketitle

Based on Floquet's theorem \cite{Floquet1883}, the stroboscopic dynamics of periodically driven quantum systems is effectively described by a time-independent Hamiltonian which can be engineered by controlling the properties of the drive \cite{Shirley1965,PhysRevX.4.031027}. This type of Floquet Hamiltonian Engineering (FHE) leads to the observation of new phenomena in quantum many-body systems including novel phases of matter \cite{Zhang2017,Choi2017.}, dynamical phase transitions \cite{Jurcevic2017}, lattice gauge theories \cite{Schweizer2019} and Floquet-topological matter \cite{Jotzu2014,Flaschner2016}. A paradigmatic example consists in a series of spin echo sequences \cite{Hahn1950,Carr1954} applied to a system of spins embedded in an inhomogeneous environment, as encountered for example as microscopic magnetic fields in nuclear magnetic resonance (NMR). The pulse sequence inverts the spins multiple times such that they finally return to their initial states and appear to be effectively decoupled from the environment. In the framework of a Hamiltonian description, this corresponds to the emergence of a vanishing Hamiltonian, i.e. the effective time-independent Hamiltonian over the spin echo sequence becomes zero. While this type of sequences is suitable to decouple the dynamics of single spins, more sophisticated sequences have been used to decouple two-body interactions \cite{Waugh1968} and to engineer Hamiltonians for studying many-body localization of mixed quantum states \cite{Wei2018}. 

The application of multi-pulse protocols, as, e.g., introduced by \cite{Choi2020,Ajoy2013}, to closed quantum systems opens up a new avenue towards quantum engineering of many-body spin systems. Due to the genuine decoupling from the environment and the high degree of control down to the single particle level, system's universality classes, symmetries and type of interactions would become tunable. Here, we demonstrate the realization of such programmable many-body spin Hamiltonians and observe the modification of the out-of-equilibrium dynamics
as a function of the target Hamiltonian's parameters. We use an ultracold atomic Rydberg gas which can be well decoupled from the environment and has already been established as an excellent platform for quantum simulation of closed systems \cite{Gross995,Browaeys2020} enabling, e.g., the preparation of pure many-body quantum states in random \cite{Orioli2018} and controllable spatial geometries  \cite{Bernien2017,Labuhn2016}. The unitary dynamics of this system is captured by spin models \cite{Adams2015}, which serve as prototypical templates for studying quantum magnetism emerging from different classes of Hamiltonians \cite{Dmitriev_2002}.

The key idea of this work is to transform the naturally given spin Hamiltonian $H_{\rm{Nat}}$ into a target Floquet Hamiltonian $H_{\rm{Floq}}$ using a sequence of periodically applied control pulses (Fig.~1A). To demonstrate programming of the resulting effective Hamiltonian we engineer a tunable Heisenberg XYZ Hamiltonian from a naturally given XX Hamiltonian. We benchmark FHE by comparing the emerging dynamics of the driven system to the one expected from an effective time-independent Hamiltonian. As a demonstration of the efficient tunability, we modify the symmetries in an effective XYZ Heisenberg model and analyze the resulting change of the out-of-equilibrium relaxation dynamics.


More specifically, we consider a Heisenberg XX Hamiltonian with an external driving field
\begin{equation}
    H(t) = H_{\rm{XX}} + H_{\rm{drive}}(t),
\label{Eq.H(t)}
\end{equation}
where $H_{\rm{XX}} = \sum_{i,j} J_{ij}/\hbar  \left(S_{x}^{i}S_{x}^{j} + S_{y}^{i}S_{y}^{j} \right)$  and $H_{\rm{drive}}(t) = \sum_{i} \Omega(t) \left[ \text{cos}\phi (t) S_{x}^{i} + \text{sin}\phi (t) S_{y}^{i} \right]$. Here, $S_{\alpha}^{i}$ ($\alpha \in {x,y,z}$) are spin-1/2 operators, $\Omega(t)$ the time-dependent Rabi frequency of the drive, $\phi(t)$ the phase and $J_{ij}$ the interaction coefficient between spin $i$ and $j$. As schematically depicted in Fig. 1B, we drive the system with a periodic sequence of four global $\pi/2$ pulses realizing different spin operator ($S_x,-S_y,S_y,-S_x$) in $H_{\rm{drive}}(t)$ followed by delay times $\tau_1 = \tau (1-2v+2w)$ , $\tau_2 = \tau (1+2u-2w)$ and $\tau_3 = \tau (1-2u+2v)$, where $u,v,w$ are dimensionless parameters. Under the periodic driving $H_{\rm{drive}}(t)$ that rotates the spin frame, a time-independent Hamiltonian $H_{\rm{Floq}}$ can be associated to $H(t)$ within the framework of \onecolumngrid\
\begin{figure}[t]
\includegraphics[width=0.66\textwidth]{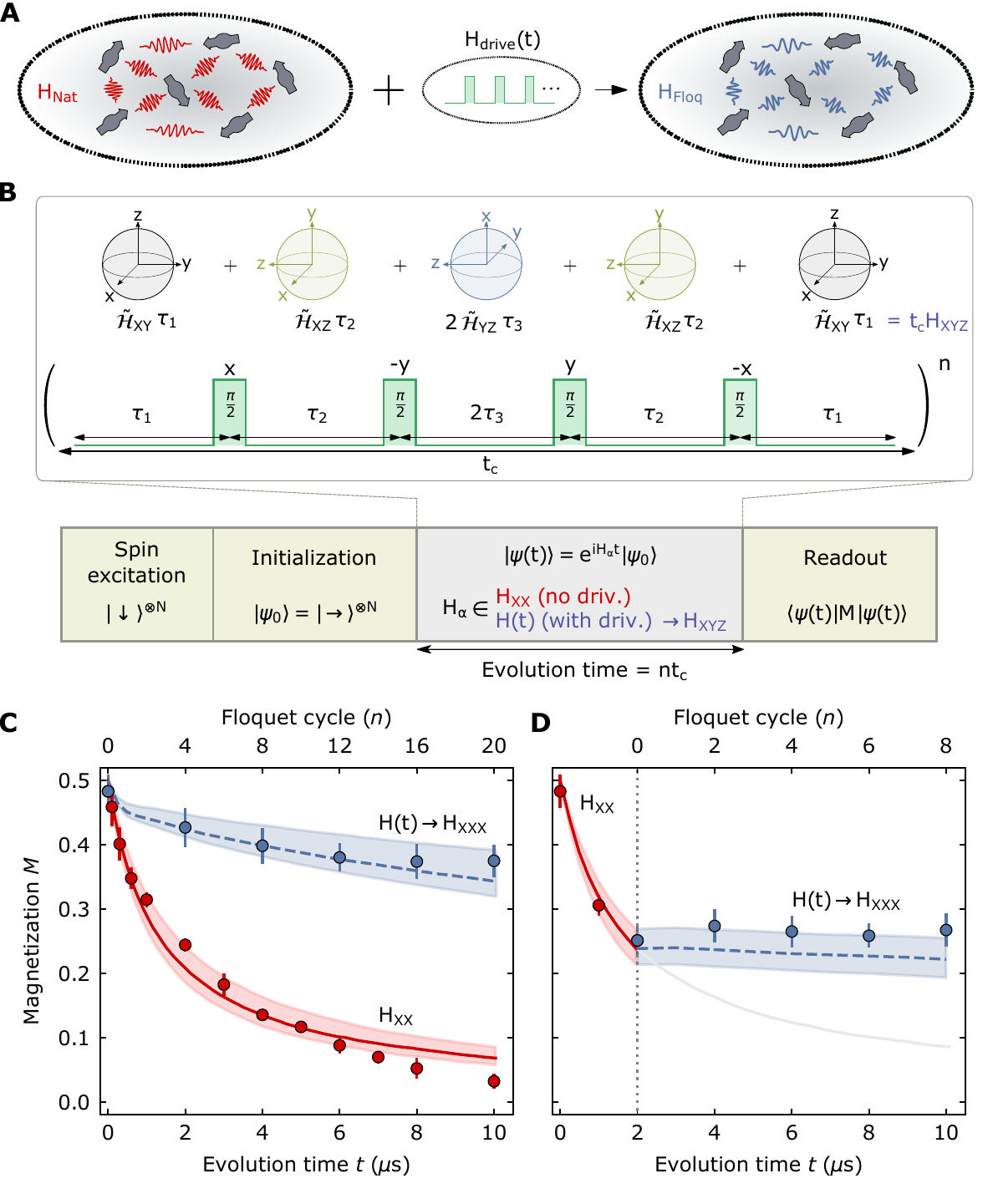}
\caption{\textbf{Hamiltonian Engineering of an isolated spin system} \textbf{(A)} A periodically applied Hamiltonian $H_{\rm{drive}}(t)$ is used to transform the natural Hamiltonian $H_{\rm{Nat}}$ of the system into a time-independent Floquet Hamiltonian $H_{\rm{Floq}}$. In this work we have chosen $H_{\rm{Nat}}$ to be represented by a Heisenberg XX Hamiltonian $H_{\rm{XX}}$. \textbf{(B)} Experimental protocol for measuring the total magnetization of the system under periodic driving composed by $n$ repetitions of four $\pi/2$ pulses in different directions ($x,-y,y,-x$) with adjustable delay time. The spheres indicate the spin-frame transformations with the resulting piece-wise constant rotated-frame Hamiltonians $\tilde{\mathcal{H}}_{\alpha \beta} = \sum_{i,j} J_{ij}/\hbar  \left(\tilde{S}_{\alpha}^{i}\tilde{S}_{\alpha}^{j} + \tilde{S}_{\beta}^{i}\tilde{S}_{\beta}^{j} \right)$ below. \textbf{(C)} Magnetization dynamics obtained by the natural Hamiltonian $H_{\rm{XX}}$ (red points) and by the driven Hamiltonian $H(t)$ for $t_c =0.5~\mu\rm{s}$ (blue points) and $J_{\text{m}}/2\pi=0.2$ MHz. \textbf{(D)} Magnetization dynamics when the Floquet cycles start after $2~\mu\rm{s}$ for $t_c =1~\mu\rm{s}$ and $J_{\text{m}}/2\pi=0.2$ MHz. Lines are dTWA simulations of the respective dynamics under $H_{\rm{XX}}$ (solid) and  $H(t)$ (dashed) including uncertainties of the Rydberg density (shaded area). Error bars are the statistical errors resulting from 100 repetitions of the experiment.}
\label{fig:Fig1}
\end{figure}
\twocolumngrid\ Average Hamiltonian Theory (AHT) \cite{SeeOnline}. Each $i_\text{th}$ pulse is rotating the spin-frame such that the system evolves under the action of a rotated-frame Hamiltonian $\tilde{\mathcal{H}}_i$ during the 
propagation time $\tau_i$ (Fig. 1B top) \cite{Choi2020}; a thorough derivation can be found in the supplementary material \cite{SeeOnline}. After an integer number of Floquet cycles, the resulting zeroth order Floquet Hamiltonian is $H_{\rm{Floq}} = \frac{1}{t_c} \sum_{i=1}^5 \tilde{\mathcal{H}}_i \tau_i$ and, for our specific implementation, reads
\begin{equation}
H_{\rm{XYZ}} =  \frac{2}{3}\sum_{i,j}  J_{ij}/\hbar \left( \delta_x S_{x}^{i}S_{x}^{j} + \delta_y S_{y}^{i}S_{y}^{j}
 + \delta_z S_{z}^{i}S_{z}^{j}  \right) \quad ,
\label{eq: Hxyz}
\end{equation}
where $ \delta_x = 1-v+u$, $\delta_y = 1+w-u$ and $\delta_z = 1-w+v$ are the  control parameters allowing one to transform the natural XX Hamiltonian into an effective XYZ form. According to AHT, this description is valid as long as the period of the pulse sequence i.e. the cycling time $t_c$ is much shorter than the typical time scales of the mutual interactions $J_{ij}$, as expressed by the condition 
\begin{equation}
   J_{ij}  \cdot t_c \ll 2\pi.
\label{eq: JtCrit}
\end{equation}

In our experimental implementation, the spin is represented by the two Rydberg states $\ket{48S_{1/2},m_j = 1/2}=\ket{\downarrow}$ and $\ket{48P_{3/2},m_j = 1/2}=\ket{\uparrow}$ in an ultracold gas of rubidium atoms. The two states couple via dipole-dipole interaction yielding coefficients $J_{ij}=2\,C_3(\theta_{ij})/r_{ij}^3$ in Eq.~\ref{Eq.H(t)}, with the angle-dependent dipolar coupling parameter $C_3(\theta_{ij})$ and $r_{ij}$ the spatial separation between atom $i$ and $j$ \cite{SeeOnline,Browaeys2020}. The system of a few hundred spins is disordered due to the random positions of the $N$ Rydberg atoms in the cloud. The mean value of the interaction strength, defined as $J_\text{m}=1/N \sum_i \vert \sum_j J_{ij} \vert$, can be modified by changing the Rydberg density. A microwave field resonantly coupling the two Rydberg states allows us to globally manipulate the spins. The field is controlled by an Arbitrary Waveform Generator (AWG) with adjustable amplitude and phase \cite{SeeOnline}. The system's total magnetization $M$ is mapped out by a tomographic measurement \cite{SeeOnline}.

In a first series of experiments, we choose the pulse sequence such that $u=v=w$, which is formally equivalent to the so-called WAHUHA sequence~\cite{Waugh1968} widely used in NMR to realize dynamical decoupling and to suppress spin-spin interactions.  For the isolated quantum system considered here, this particular sequence implies  $\delta_x=\delta_y=\delta_z$ resulting in the engineering of a symmetric Floquet Hamiltonian $H_{\rm{Floq}} = H_{\rm{XXX}}$, where the total spin constitutes a conserved quantity \cite{SeeOnline}.  

The experimental protocol, as shown in Fig.~\ref{fig:Fig1}B, starts with the excitation of the $\ket{\downarrow}^{\otimes N}$ Rydberg state, followed by a $\pi/2$ pulse, thus initializing the spins in the product state $\ket{\rightarrow}_x^{\otimes N} = 1/\sqrt{2}(\ket{\downarrow}+\ket{\uparrow}  )^{\otimes N}$ (total magnetization pointing along the $x$-direction). 
The spin system then evolves under the Hamiltonian given by Eq.~\ref{Eq.H(t)}. After time $t$, the total magnetization is measured.

Without periodic driving the dynamics is governed by $H_{\rm{XX}}$ resulting in a fast relaxation towards a demagnetized state within $\sim10~\mu$s (red points in Fig. 1C). In stark contrast, we observe a significant slowing down of the relaxation dynamics when we apply the WAHUHA sequence with $H_{\rm{XXX}}$ as target Hamiltonian (blue points in Fig. 1C). To confirm that the stalling of the dynamics is independent of the initial state, we let the system first evolve under $H_{\rm{XX}}$ for $2~\mu s$ to create an entangled state. We then apply the WAHUHA sequence and find a complete freezing out of the magnetization dynamics (see Fig. 1D) as indeed expected for the symmetric $H_{\rm{XXX}}$.

\begin{figure}[t]
\includegraphics[width=0.33\textwidth]{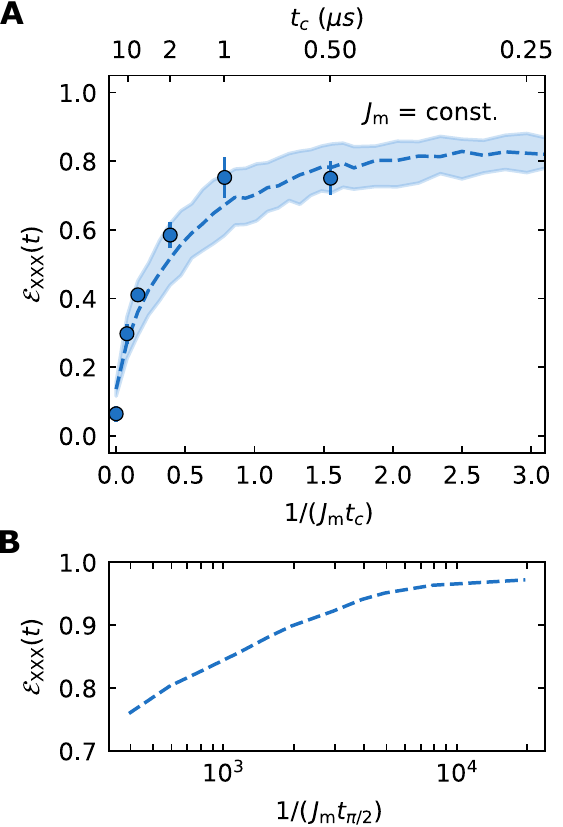}
\caption{\textbf{Efficiency of the engineering} \textbf{(A)} 
Efficiency at $t=10$ $\mu$s as function of $1/(J_\text{m} t_c)$ for a fixed mean interaction strength $J_\text{m}/2\pi=0.2$ MHz and $t_{\pi/2} = 12.5$ ns. Dashed lines are dTWA simulations of $H(t)$ with the respective sequence parameters and density uncertainties (shaded area). \textbf{(B)} Numerical dTWA simulations of the efficiency at $t=10$ $\mu$s with pulse widths ranging from $t_{\pi/2} = 12.5$ ns to $0.25$ ns for $t_c=50$ ns and $J_\text{m}/2\pi=0.2$ MHz.}
\label{fig:Fig2}
\end{figure}

To understand why the evolution in Fig. 1C is not entirely frozen we have to consider the disordered nature of the system. The distance between a pair of Rydberg atoms has a lower bound $r_b$, determined by the so-called Rydberg blockade induced by the laser excitation process \cite{Comparat2010}, resulting in a maximum interaction  strength of $J_{\rm{max}}/2\pi= 2 \, C_3/(2\pi r_{\rm{b}}^3) = 18$ MHz.  For the closest pairs, this turns out to be large compared to the inverse cycling time $1/t_c= 2$ MHz thus violating the condition Eq. \ref{eq: JtCrit} required for the validity of an effective time-independent Hamiltonian description. The result is a slow remnant relaxation of the total magnetization. In Fig. 1D, however, the strongest interacting pairs of spins have demagnetized before the driving is applied and thus no longer contribute to the total magnetization. This qualitative interpretation is confirmed by a semi-classical simulation based on the discrete Truncated Wigner Approximation method (dTWA) \cite{Schachenmayer2015} for the full time-dependent Hamiltonian given by Eq.~\ref{Eq.H(t)} which quantitatively reproduces all essential features of the magnetization dynamics.  

In order to quantify experimental deviations from ideal FHE, we introduce an efficiency for engineering the target Hamiltonian $H_{\text{XXX}}$ defined as $\mathcal{E}_\text{XXX}(t)= \frac{ M (t)}{1/2}$, where $ M (t)$ is the observed magnetization after a time $t$. 
An efficiency of $\mathcal{E}_\text{XXX}(t) \equiv 1$ corresponds to the dynamics expected for perfectly matching an XXX Hamiltonian.  Figure 2A shows the efficiency as a function of $1/(J_\text{m} t_c)$ varying $t_c$ while keeping $J_\text{m}$ fixed. If no pulse sequence is applied, which corresponds to $t_c$ reaching infinity, the efficiency is close to zero due to the relaxation of magnetization under the action of a pure $H_\text{XX}$ as shown in Fig. 1C. For shorter $t_c$ a growing number of spin pairs fulfill the condition of Eq. \ref{eq: JtCrit} resulting in an increase of the efficiency. Eventually, for the shortest achievable cycling times, the efficiency approaches a finite value in good quantitative agreement with the dTWA prediction.

The fact that the efficiency does not reach unity under our experimental conditions can be attributed to the influence of interactions during the application of the pulses. This influence would become negligible if the pulse length  $t_{\pi/2}$ was reduced beyond the current shortest value given by the bandwidth of our AWG. In Fig. 2B we determine the reachable efficiency by numerically simulating pulse lengths down to $t_{\pi/2} = 0.25$ ns at fixed $1/(J_\text{m} t_c) =  16$.
Under these conditions, which are within reach with today's cutting-edge microwave technology, the efficiency would indeed approach unity.
 
\begin{figure}[t]
\includegraphics[width=0.33\textwidth]{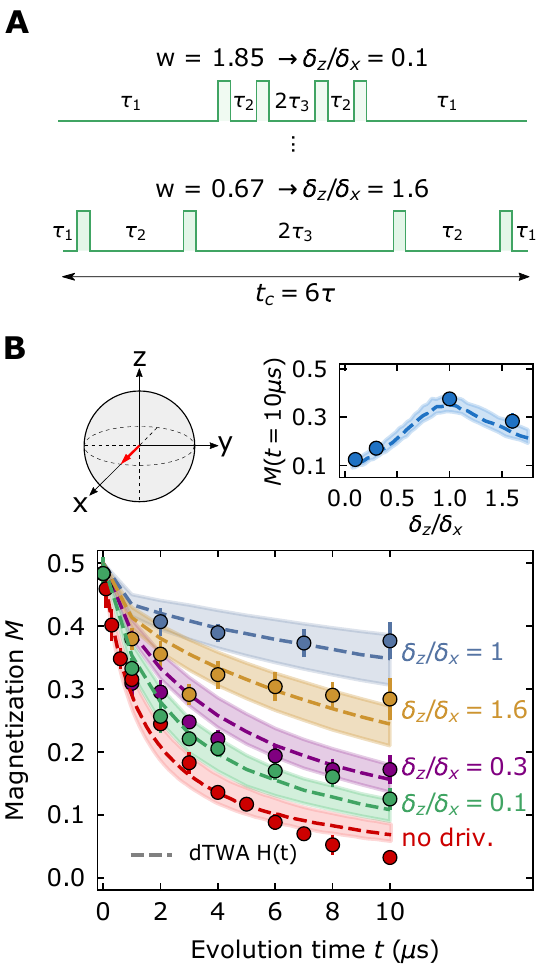}
\caption{\textbf{Engineered out-of-equilibrium dynamics} \textbf{(A)} Schematic representation of the engineering sequence for the specific case of a XXZ Hamiltonian allowing to modify the anisotropy parameter $\delta_z/\delta_x$ for fixed $u=(w+v)/2$ and $v=1$. \textbf{(B)} Magnetization dynamics for different pulse delays with $t_c = 1~\mu s$, $t_{\pi/2} = 12.5$ ns, $J_\text{m}/2\pi=0.2$ MHz. Shaded area corresponds to the density uncertainty in dTWA simulations of $H(t)$. The red arrow in the sphere is indicating the initial state.}
\label{fig:Fig3}
\end{figure}

To demonstrate the full potential of FHE for programmable quantum simulation we now proceed to a larger range of Hamiltonians possessing different symmetries and explore the resulting dynamics for different initial states. The freezing of the magnetization under the engineered XXX Hamiltonian observed in the previous set of measurements (see Figs. 1C,D) is a direct consequence of a $SU(2)$ symmetry in the system. Using the pulse sequence shown in Fig. 3A we can break this symmetry by setting $\delta_x = \delta_y \neq \delta_z$ effectively engineering a XXZ Hamiltonian which does not conserve the magnetization for a state initialized in $x$-direction. 

Figure 3B shows the time evolution of the magnetization for different values of the anisotropy parameter $\delta_z / \delta_x$  ranging from 0.1 to 1.6. Depending on the choice of $\delta_z / \delta_x$ the magnetization evolves in a non-exponential fashion typical for a disordered system consisting of hundreds of spins \cite{Signoles2019}. The noticeable feature is a non-monotonous change of the observed relaxation dynamics as a function of $\delta_z / \delta_x$,  which is well reproduced by the dTWA simulation of the time-dependent Hamiltonian (see inset) and in full agreement with the expectation for the time-independent target XXZ Hamiltonian (see \cite{SeeOnline}).

The XXZ Hamiltonian still features a $U(1)$ symmetry, which results in a conservation of the magnetization along $z-$direction. To illustrate 
that the conservation is indeed a property of the Hamiltonian and not of the initial state, we initialize the system by a $\pi/4$ pulse resulting in non-zero magnetization components along the $y-$ and $z-$direction (see Fig. 4A). As shown in Fig. 4B, the time evolution clearly differs for each of the magnetization components. In particular, the $y$-magnetization relaxes while the $z$-magnetization remains essentially constant as a consequence of the $U(1)$ symmetry. Breaking this remaining $U(1)$ symmetry, by setting the pulse sequence to $\delta_x \neq \delta_y \neq \delta_z$ (fully anisotropic XYZ Hamiltonian), induces a non-conversation of the magnetization in $z-$ direction resulting in a relaxation of all magnetization components as shown in Fig. 4C. One observes that the $z-$magnetization relaxes faster than the $y-$magnetization. This is due to the fact that the dynamics of the $z-$magnetization scales with $\delta_x$-$\delta_y=0.9$ while the dynamics of the $y-$magnetization scales with $\delta_x$-$\delta_z=0.45$  
(for more details, see \cite{SeeOnline}).


Floquet Hamiltonian Engineering of isolated spin systems opens up new possibilities for quantum simulation of many-body systems beyond equilibrium. The ability to synthesize large varieties of many-body Hamiltonians allows the experimental exploration of important concepts such as spin transport \cite{Jepsen2020}, the generalization of fluctuation dissipation relations \cite{PineiroOrioli2019} to out-of-equilibrium systems and the nature of thermalization mechanisms \cite{Rigol2008}. For disordered systems, Floquet Hamiltonian Engineering technique can be associated to the realization of time-reversal operations \cite{Wei2018}, with the aim of measuring Out-of-Time-Order Correlators (OTOCs), in order to experimentally investigate the fundamental role of entanglement and information scrambling in out-of-equilibrium systems \cite{Garttner2017,Landsman2019} as well as the existence of non ergodic many-body localized states \cite{Smith2016} in various spin Hamiltonians. The approach can directly be combined with current developments, e.g. in preparing Rydberg atom arrays \cite{Browaeys2020}, thus combining efficient engineering of the Hamiltonian with precise control over the spatial arrangement provides a viable scenario for the realization of fully programmable quantum spin simulators.

\vspace{0.7mm}
The authors gratefully acknowledge insightful discussions with M. Gärttner, S. Whitlock, P. Cappellaro and K. X. Wei. This work has been supported by the by the DFG under Germany’s Excellence Strategy EXC 2181/1 - 390900948 (the Heidelberg STRUCTURES Excellence Cluster) and the European Commission FET flagship project PASQuanS (Grant No. 817482). It is part of the DFG Collaborative Research Centre SFB 1225 (ISOQUANT) and the DFG Priority Program 1929 “GiRyd” (DFG WE2661/12-1). Support by the Heidelberg Center for Quantum Dynamics is gratefully acknowledged. N.T. received funding from the European Union’s Horizon 2020 program under Marie Sklodowska-Curie grant agreement no. 798402. C.H. is supported by the Alexander von Humboldt foundation. T.F. receired a graduate scholarship of the Heidelberg University (LGFG).

\onecolumngrid\
\begin{figure}[t]
\includegraphics[width=0.66\textwidth]{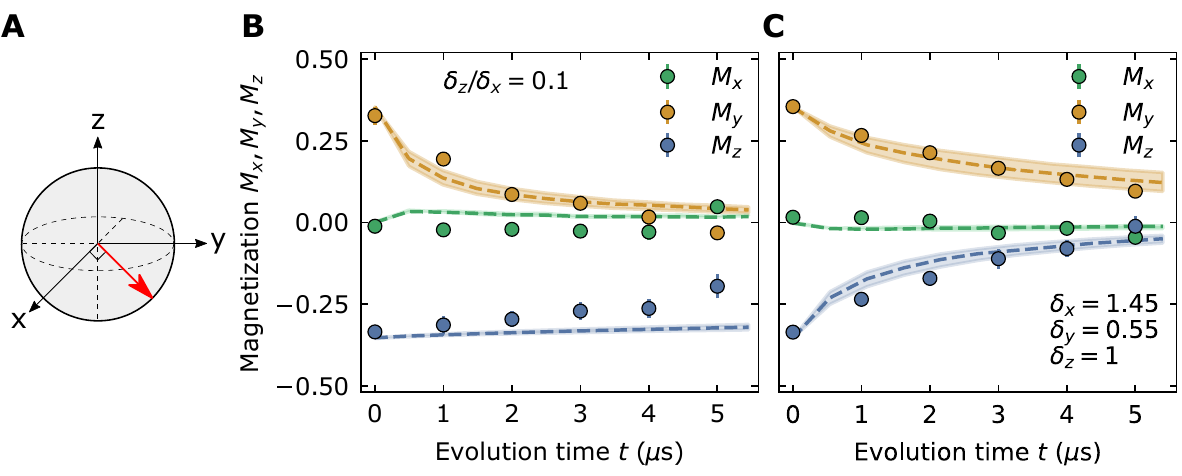}
\caption{\textbf{Consecutive symmetry breaking for a dual component initial state} \textbf{(A)} Representation of the state initialized with a $\pi/4$ pulse on the Bloch sphere. \textbf{(B)} Dynamics of the magnetization components for a XXZ Hamiltonian. \textbf{(C)} Dynamics of the magnetization component for a fully anisotropic XYZ Hamiltonian. Parameters are $t_c = 0.5~\mu s$, $t_{\pi/2} = 10.7$ ns, $J_\text{m}/2\pi=0.4$ MHz and the shaded area corresponds to the density uncertainty of dTWA simulations of $H(t)$ (dashed lines).}
\label{fig:Fig4}
\end{figure}
\twocolumngrid\
\bibliography{Arxbib}

\clearpage
\onecolumngrid

\section*{Supplementary Materials}

\section*{Methods}

\subsection*{State preparation}

The preparation of the spin-$1/2$ system starts with a cloud of ultracold Rubidium 87 atoms in an optical dipole trap, which density distribution is described by a Gaussian distribution (radii at $1/e^2$ : $\sigma_x = 397~\mu \rm{m}$, $\sigma_y = \sigma_z = 313~\mu \rm{m}$). With a temperature of $T \sim 50 \mu K $, we expect the atoms motional degree of freedom to be frozen over the experimental time scales. Using optical pumping techniques and microwave Landau-Zener passages, the atoms are prepared in the $\ket{g} = \ket{5S_{1/2},F=2,m_\text{F}=2}$ ground state from where they are excited to the $\ket{48S_{1/2},m_j = 1/2}$=$\ket{\downarrow}$ Rydberg state via a two-photon laser excitation at wavelengths of $780$\,nm and $480$\, nm with a single photon detuning of $\Delta/2\pi = 97 $ MHz from the intermediate state $\ket{e} = \ket{5P_{3/2},F=3,m_\text{F}=3}$. The two beams have respective waists of $3000$\, $\mu\rm{m}$ ($780$\, nm) and $268\,$ $\mu\rm{m}$ ($480$\,nm) and are aligned in a counter-propagating configuration leading to an effective two-photon coupling of $\Omega_\text{eff}/2\pi = 15$ kHz. Excitation pulses with variable length allow the excitation of up to $1000$ Rydberg spins. To address individual states in the Zeeman manifold, we apply a magnetic field of $B = 30 $\, G.
Using a microwave field we couple the spin state $\ket{\downarrow}$ to the $\ket{48P_{3/2},m_j = 1/2}$=$\ket{\uparrow}$ state.
The resonant dipole interaction between Rydberg atoms in these two states can be described by the Heisenberg XX Spin model Eq.~\ref{Eq.H(t)} with interaction strength $J_{ij}$ where $J_{ij} =2\, C_3(1-3\text{cos}^2\theta_{ij})/r_{ij}^3$. Here, $C_3/2\pi = 1.14~\text{GHz} \cdot \mu \rm{m^3}$ is the coefficient of the dipolar exchange interaction . To ensure unitary Hamiltonian dynamics, we restrict the experimental time scales to a maximum of $10$ $\mu\rm{s}$ which is short compared to the spontaneous decay time ($113$ $\mu\rm{s}$) and the redistribution time ($121$ $\mu\rm{s}$) to other Rydberg states induced by black-body radiation \cite{Sibalic2017}.

\subsection*{Distribution of interaction strengths}

To model the experimental 3D spin distribution, we simulate the Rydberg excitation dynamics \cite{Signoles2019} in a cloud of ground state atoms with the experimentally given parameter.  The dephasing rate of the two-photon transition is adjusted to $\gamma/2\pi=110~\rm{kHz}$ such that the total number of excited atoms equals the one measured by the calibrated field ionization. The obtained 3D spin distribution allows us to deduce the distribution of interaction strengths $J_{ij}$. From this we calculate the mean interaction strength $J_{\text{MF}}^{i}=\sum_{j} J_{ij}$ for each spin $i$ and deduce the  mean interaction strength $J_{\rm{m}}$ of the system by averaging the absolute value of the resulting $J_{\text{MF}}^{i}$ distribution: $J_\text{m}=1/N \sum_i \vert J_{\text{MF}}^{i} \vert$.

\subsection*{Microwave control}

The manipulation of the spins is realized with a microwave field that couples $\ket{48S_{1/2},m_j = 1/2}$ to $\ket{48P_{3/2},m_j = 1/2}$ resonantly with a frequency of $\nu = 35.5 $ GHz that acts as an additional external field on the spin Hamiltonian described by
\begin{equation}
    H_\text{drive}(t) = \sum_{i} \Omega(t) ( \text{cos}\phi (t) S_{x}^{i} + \text{sin}\phi (t) S_{y}^{i} ) .
    \label{eq: H_field}
\end{equation}
Amplitude and phase of the field can be tuned
with an AWG (Tabor SE$5082$, sampling rate of $5$ Gsample per second, analog BW of $1.2$ GHz) at a frequency of $\nu \sim 800 $ MHz. The wave is then mixed with a $\nu = 34.7$ GHz local oscillator (Anritsu MG$3697$) by frequency up-conversion scheme using IQ mixing. The field is emitted by a horn antenna (SGH-$22$) and passes through a wire grid polarizer to ensure a well defined linear polarization. We determine that 97$\%$ of the field couples to the desired $\pi-$transition between $\ket{48S_{1/2},m_j = 1/2}$ and $\ket{48P_{3/2},m_j = 1/2}$. Finally, a parabolic aluminum mirror is used to focus the microwave radiation to the atoms down to a waist of $\sigma_{\rm{mw}} = 20$ mm.

\begin{figure*}[t]
\includegraphics[width=0.66\textwidth]{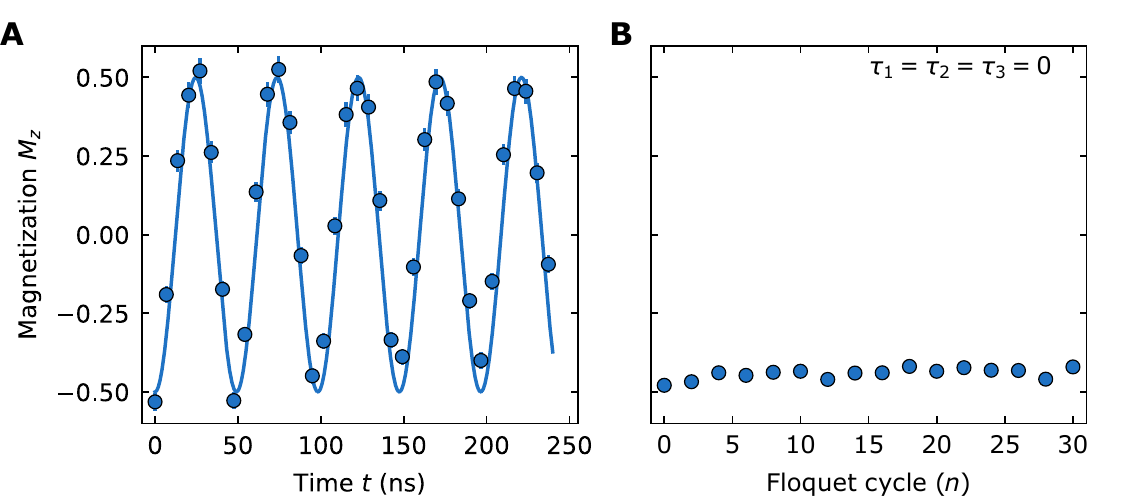}
\caption{\textbf{Microwave characterization.} \textbf{(A)} Rabi oscillations between $\ket{48S_{1/2},m_j = 1/2}$ and $\ket{48P_{3/2},m_j = 1/2}$ states with  $J_\text{m}/2\pi=0.2$ MHz. Solid line corresponds to a cosine fit with the Rabi frequency $\Omega$ as a free fitting parameter. \textbf{(B)} Revival of the z-magnetization after the sequence of Fig~\ref{fig:Fig1}B with no delay between the pulses for variable number of Floquet cycles and thus $\pi/2$ pulses. Error bars are determined from the standard error of the mean value.}
\label{fig:FigSM}
\end{figure*}

To characterize the microwave setup we perform Rabi oscillations and spin echo experiments. Figure~\ref{fig:FigSM}A displays Rabi oscillations of the magnetization in $z$-direction $M_z$ between $\ket{\downarrow}$ and $\ket{\uparrow}$ with a frequency of $\Omega/2\pi = $ $20.33(7)$ MHz. To verify that our microwave driving is not adding significant additional technical noise that might effect the Hamiltonian engineering, we perform the four-pulse sequence presented in the main text without time delay between the pulses ($\tau_1 = \tau_2 = \tau_3 = 0$) and measure $M_z$. By the choice of the pulses, the spin rotations within on cycle of the form $(x, -y, y, -x)$ cancel each other such that we expect a preservation of the initial state. The sequence is applied directly after the Rydberg excitation ($\ket{\downarrow}^{\otimes N}$) and repeated for $n$ times. Figure~\ref{fig:FigSM}B shows that even after $30$ cycles, which corresponds to $120$ $\pi/2$ pulses, $93.5 \%$ of the magnetization remains in the z-direction. This measurement indicates that within the $ 4$ to $80$ pulses applied in the experiments presented in the main text, we have less than $6.5 \%$ accumulated error in the rotation angle of the spins.

\subsection*{Determination of the magnetization}

The determination of the magnetization is performed by measuring the population in different Rydberg states using electric field ionization and subsequent ion detection \cite{Signoles2019}. The ion detector is calibrated using the method of depletion imaging \cite{Ferreira_Cao_2020} of Rydberg atoms from which we deduce a detection efficiency of $\eta = 0.102$ ions per Rydberg atom. From the measurement of the Rydberg population and tomographic readout \cite{Signoles2019}  we determine the ensemble averaged spin components ($M_x = 1/(\hbar N) \sum_{i}^{N} \langle S_x^{i}\rangle$, $M_y = 1/(\hbar N) \sum_{i}^{N} \langle S_y^{i}\rangle$, $M_z = 1/(\hbar N) \sum_{i}^{N} \langle S_z^{i}\rangle$) from which we calculate the total magnetization $ M = \sqrt{M_x^2+M_y^2+M_z^2}$.

\section*{Supplementary Text}

\subsection*{Average Hamiltonian Theory and derivation of the Floquet Hamiltonian}

The derivation of the time-independent target Hamiltonian from the periodically driven time-dependent Hamiltonian $H(t) = H_{\rm{XX}} + H_{\rm{drive}}(t)$ can be understood in in the framework of Average Hamiltonian Theory (AHT) \cite{Haeberlen,Choi2020}.

The evolution of the driven spin systems density matrix is captured by $\dot{\rho} = -i[H(t),\rho]$. The interaction picture evolution operator defined by $U_{\rm{drive}}(t) = \mathcal{T} \rm{exp}[-i \int_{0}^{t}H_{\rm{drive}}(t^\prime)dt^\prime]$ ($\mathcal{T}$ is the time ordering operator),  transforms the system into a rotated-frame given by $\tilde{\rho} = U^\dagger_{\rm{drive}}(t) \rho U_{\rm{drive}}(t)$, and by the rotated-frame Hamiltoninan  $\tilde{\mathcal{H}}(t) = U^\dagger_{\rm{drive}}(t) H_{\rm{XX}}U_{\rm{drive}}(t)$. For periodically applied sequences the evolution operator $U_{\rm{drive}}(t)$ and thus $\tilde{H}(t)$ is periodic in time with periodicity $t_c$. The evolution operator over one cycle is chosen to be identity  $U_{\rm{drive}}(t_c) =\mathbb{1}$ such that the systems dynamic in both frames ($\rho$ and $\tilde{\rho}$) is identical for stroboscopic observations at multiple integers of the cycling time $ t=n t_c$. Using AHT, the systems evolution operator over one cycle can be written as $U(t_c) = \exp{[-i H_{\rm{Floq}} t_c]}$, where $H_{\rm{Floq}}$ is the Floquet Hamiltonian (also known as Average Hamiltonian) and thus the systems appears to evolve under a time-independent Hamiltonian. 
We obtain $H_{\rm{Floq}}$ using a Magnus expansion by writing $H_{\rm{Floq}} =H^{(0)}+H^{(1)}+...$ . The first few terms are given by
\begin{equation}
\begin{split}
    H^{(0)} &= \frac{1}{t_c}\int_{0}^{t_c} \tilde{\mathcal{H}}(t') dt' \\
    H^{(1)} &= \frac{-i}{2t_c}\int_{0}^{t_c}dt' \int_{0}^{t'}dt [\tilde{\mathcal{H}}(t'),\tilde{\mathcal{H}}(t)].
    \label{Eq.Magnus}
\end{split}    
\end{equation}
The accuracy of the Floquet Hamiltonian depends on the truncation order of the expansion. For cycling times $t_c$ that are much faster than the typical time scales of system Hamiltonian $H_{\rm{XX}}$ (given by the interactions), the first order is typically a good approximation for $H_{\rm{Floq}}$. Additionally, the pulse sequence used in the main text is chosen to be symmetric such that all the odd orders of the expansion (including $H^{(1)}$) become zero and we thus focus on the zeroth order contribution of the Floquet Hamiltonian.

The four pulse sequence of the main text contains four $\pi/2$ pulses denoted by $(x, -y, y, -x)$ with respective phases $\phi$ (see Eq.~\ref{eq: H_field}) of ($0, 3\pi/2, \pi/2, \pi$), which are followed by delay times $\tau_1, \tau_2$ and $2\tau_3$. Assuming delta-like pulses, the integral in Eq. 5 reduces to a sum over 5 piece-wise constant rotated-frame Hamiltonian $\tilde{\mathcal{H}}_i$ weighted by their corresponding propagation time $\tau_i$ : 
\begin{equation}
 H^{(0)} = H_{\rm{Floq}} = \frac{1}{t_c} \sum_{i=1}^5 \tilde{\mathcal{H}}_i \tau_i.
\end{equation}

In order to provide a simple example of a typical AHT calculation of $\tilde{\mathcal{H}}_i$, we consider the particular case of a two-particle XX Hamiltonian $H_{\rm{XX}} = J/\hbar  \left(S_{x}^{1}S_{x}^{2} + S_{y}^{1}S_{y}^{2} \right)$. 
Before the first $\pi/2$ pulse (when the rotated-frame and lab-frame are the same) we acquire a component $\tilde{\mathcal{H}}_1 \tau_1/t_c=  J/\hbar (\tilde{S}_{x}^{1}\tilde{S}_{x}^{2}+\tilde{S}_{y}^{1}\tilde{S}_{y}^{2})\tau_1/t_c$. The first $\pi/2$ pulse in $x$-direction transform the spin frame (see spheres in Fig. 1B) and rotates $S_y \rightarrow U_{\pi/2_x}^\dagger S_y U_{\pi/2_x} = \tilde{S}_z$ while $S_x$ remains unchanged. Thus, we acquire a component $\tilde{\mathcal{H}}_2 \tau_2/t_c=  J/\hbar (\tilde{S}_{x}^{1}\tilde{S}_{x}^{2}+\tilde{S}_{z}^{1}\tilde{S}_{z}^{2})\tau_2/t_c$ over the second evolution time $\tau_2$. Repeating this transformation over the sequence, we calculate the contributions summarized in Table 1 and derive the Floquet Hamiltonian for two particle: 
\\
$H_{\rm{Floq}} = H_{\rm{XYZ}} = \frac{1}{2(\tau_{1}+\tau_{2}+\tau_{3})}J/\hbar \left[
2(\tau_{1}+\tau_{2})\tilde{S}_{x}^{1}\tilde{S}_{x}^{2} + 2(\tau_{1}+\tau_{3})\tilde{S}_{y}^{1}\tilde{S}_{y}^{2}+2(\tau_{2}+\tau_{3})\tilde{S}_{z}^{1}\tilde{S}_{z}^{2}\right].$

\begin{table}[h!]
\centering
\begin{tabular}{||c || c c c c c c c c c||} 
 \hline
 $H_{\rm{XX}}\hbar/J$ & $\tau_1$ & $\pi/2_x$ & $\tau_2$ & $\pi/2_{-y}$ & $2\tau_3$ &  $\pi/2_{y}$ & $\tau_2$ & $\pi/2_x$ & $\tau_1$\\ [1ex] 
 \hline\hline
 $S_{x}^{1}S_{x}^{2}$ & $\tilde{S}_{x}^{1}\tilde{S}_{x}^{2}$ & $\rightarrow$ & $\tilde{S}_{x}^{1}\tilde{S}_{x}^{2}$ & $\rightarrow$ & $\tilde{S}_{y}^{1}\tilde{S}_{y}^{2}$ & $\rightarrow$ & $\tilde{S}_{x}^{1}\tilde{S}_{x}^{2}$ & $\rightarrow$ & $\tilde{S}_{x}^{1}\tilde{S}_{x}^{2}$ \\
 
 + &  &  &  &  & & & & & \\
 
 $S_{y}^{1}S_{y}^{2}$ & $\tilde{S}_{y}^{1}\tilde{S}_{y}^{2}$ & $\rightarrow$ & $\tilde{S}_{z}^{1}\tilde{S}_{z}^{2}$ & $\rightarrow$ & $\tilde{S}_{z}^{1}\tilde{S}_{z}^{2}$ & $\rightarrow$ & $\tilde{S}_{z}^{1}\tilde{S}_{z}^{2}$ & $\rightarrow$ & $\tilde{S}_{y}^{1}\tilde{S}_{y}^{2}$ \\ [1ex]
 \hline
\end{tabular}
\caption{Effect of the $\pi/2$ pulses on a two-particle Hamiltonian. The first column display the natural XX Hamiltonian with its two components. The two lower lines display how these components get rotated under the spin frame transformation by the four $\pi/2$ pulses.}
\label{table:1}
\end{table}
As the Floquet Hamiltonian only describe the stroboscopic physics of the system after each period of the driving, when the lab-frame and the rotated-frame coincide, we can use the following relation for writing the Floquet Hamiltonian: ${S}_{x}=\tilde{S}_{x},{S}_{y}=\tilde{S}_{y},{S}_{z}=\tilde{S}_{z}$. Furthermore, the calculation can be extended to more particles and the original XX Hamiltonian from the main text get transformed into a Floquet Hamiltonian given by
\begin{equation}
\begin{split}
H_{\rm{Floq}} = &\frac{1}{2(\tau_{1}+\tau_{2}+\tau_{3})}\sum_{i, j}J_{ij}/\hbar \left[
2(\tau_{1}+\tau_{2})S_{x}^{i}S_{x}^{j} + 2(\tau_{1}+\tau_{3})S_{y}^{i}S_{y}^{j}+2(\tau_{2}+\tau_{3})S_{z}^{i}S_{z}^{j}\right].
\end{split}
\label{Eq.HAHT}
\end{equation}

To obtain Eq.~2 of the main text, we rearrange the above equation by choosing $\tau_1 = \tau (1-2v+2w)$ , $\tau_2 = \tau (1+2u-2w)$, and $\tau_3 = \tau (1-2u+2v)$, where $u, v, w$ are dimensionless parameters which can then be tuned  without changing the cycling time $t_c = 2 \cdot (\tau_1 + \tau_2 + \tau_3) = 6\tau$ in the experiment.

\subsection*{Relaxation of the magnetization under the action of an XYZ Hamiltonian}

We derive the time evolution of the magnetization using the Heisenberg equation of motion for the general case of an XYZ Hamiltonian as given by Eq. 2 in the main text. Considering $S_z=\sum_{i}S_z^{i}$, $S_x=\sum_{i}S_x^{i}$, $S_y=\sum_{i}S_y^{i}$ we obtain

\begin{itemize}
    \item $dS_x/dt = i/\hbar [S_x,H_{\rm{XYZ}}] =  \left(\delta_z-\delta_y\right)\left[3\sum_{i,j,i\neq j}J_{ij}\left(S_{z}^{i}S_{y}^{j} + S_{z}^{j}S_{y}^{i}\right)/(2\hbar^2)\right]$
    \item $dS_y/dt = i/\hbar [S_y,H_{\rm{XYZ}}] =  \left(\delta_x-\delta_z\right)\left[3\sum_{i,j,i\neq j}J_{ij}\left(S_{x}^{i}S_{z}^{j} + S_{x}^{j}S_{z}^{i}\right)/(2\hbar^2)\right]$
    \item $dS_z/dt = i/\hbar [S_z,H_{\rm{XYZ}}] =  -\left(\delta_x-\delta_y\right)\left[3\sum_{i,j,i\neq j}J_{ij}\left(S_{y}^{i}S_{x}^{j} + S_{y}^{j}S_{x}^{i}\right)/(2\hbar^2)\right]$
\end{itemize}

In the scenario of the engineering of an XXX Hamiltonian, $\delta_y=\delta_x=\delta_z$ which implies $[S_x,H_{\rm{XYZ}}]=0$,  $[S_y,H_{\rm{XYZ}}]=0$ and  $[S_z,H_{\rm{XYZ}}]=0$. All the components $S_x$, $S_y$, $S_z$ are constant of motions of the system associated to the presence of the $SU(2)$ symmetry. Thus, the total magnetization constitutes a conserved quantity. For an XXZ Hamiltonian, $\delta_y=\delta_x$ which implies $[S_z,H_{\rm{XYZ}}]=0$. $S_z$ constitutes then a constant of motion which is associated to the presence of the $U(1)$ symmetry in the system. For an XYZ Hamiltonian ($\delta_x \neq \delta_y \neq \delta_z$), there is no constant of motion link to the fast that all the unitary symmetries are broken. More specifically, the dynamics of a component $S_x$ for example scales with the difference between $\delta_z$ and $\delta_y$. Qualitatively, the dynamics of $S_x$ will be faster if the value $\delta_z-\delta_y$ is important. This reasoning can be applied to all the components and explains the faster relaxation observed for the $z-$component compared to $y-$component in Fig. 4C. 

\subsection*{Numerical simulations and first-order correction terms}

We have explained in the main text that the inefficiency of the engineering observed in the experimental measurements is a consequence of too strong interactions creating additional dynamics during the application of the pulses as well as rendering the condition of Eq. 3 not valid for each spin present in the disordered sample. We propose here to investigate numerically a scenario (out-of-reach for our current experimental setup) where pulse length and cycling time can be rendered really small ($t_{\pi/2}=1.25$ ns and $t_c=50$ ns).

\begin{figure*}[t]
\includegraphics[width=0.66\textwidth]{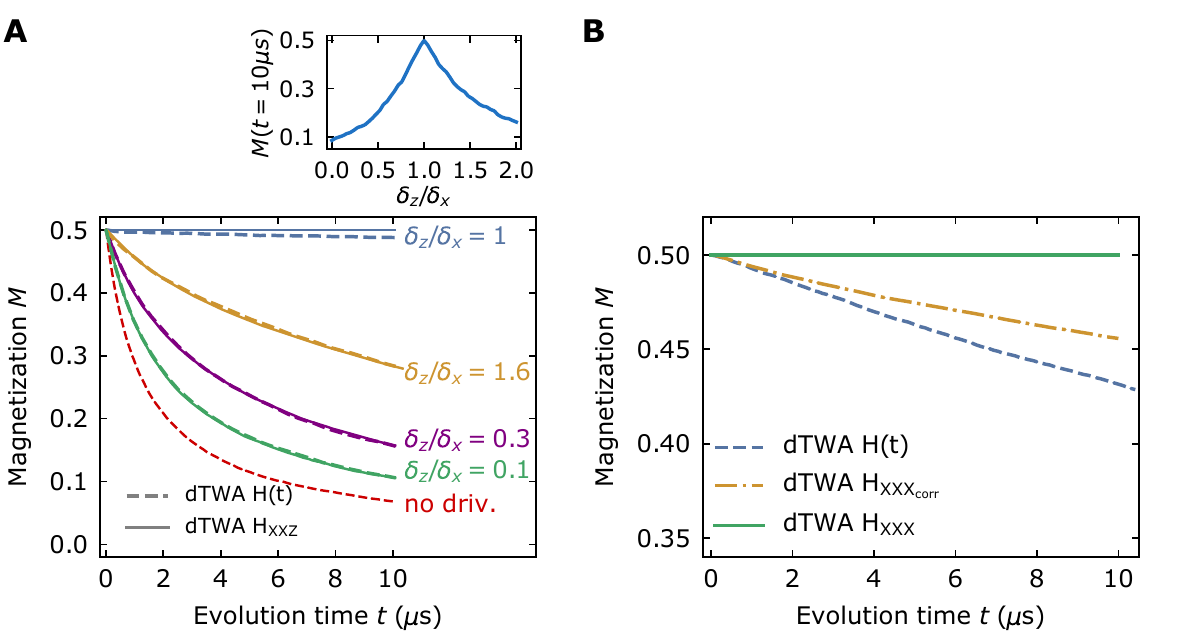}
\caption{\textbf{Numerical comparison between time-dependent and time-independent Hamiltonian} \textbf{(A)} Numerical simulation of the time-dependent Hamiltonian $H(t)$ and the anisotropic $H_{\text{XXZ}}$ Hamiltonian for various ratio $\delta_z/\delta_x$ with $J_\text{m}/2\pi=0.2$ MHz. Inset : Evolution of the magnetization at 10 $\mu$s for different ratio $\delta_z/\delta_x$ in the case of $H_{\text{XXZ}}$. \textbf{(B)} Numerical simulation of the time-dependent Hamiltonian $H(t)$, the isotropic $H_{\text{XXX}}$ Hamiltonian and the isotropic $H_{\text{XXX,corr}}$ Hamiltonian with the correction cross term with $J_\text{m}/2\pi=0.4$ MHz.}
\label{fig:FigSnumerics}
\end{figure*}

Following the experimental investigation of an anisotropic Heisenberg model in Fig. 3B, we simulate in Fig.~\ref{fig:FigSnumerics}A both the exact time dependent Hamiltonian $H(t)$ as well as the engineered time-independent Hamiltonian $H_{\text{XXZ}}$  for various ratio $\delta_z/\delta_x$. Firstly, we report that the non-monotonous dynamics observed in the experiment is well reproduced by both numerical simulations. Secondly, we see that both numerical simulations agree well within each other meaning that in this regime, $H_{\text{XXZ}}$ constitutes a really good description of $H(t)$ and the engineering can be considered reliable. 

Nevertheless, a slight discrepancy is still observed (specifically for the case of $\delta_z/\delta_x = 1$). The latter can mainly be attributed to the fact that the pulses have a finite length. In order to show precisely this point, we calculate the leading correction due to the existence of finite pulse length following the derivation performed in the reference \cite{Choi2020}. For the particular sequence of four pulses used in the main text, we obtain the following correction $\delta H_{av}$ such that : $H_{\text{XXX,corr}}=H_{\text{XXX}}+\delta H_{av}$ where
\begin{equation}
\delta H_{av}=-\frac{4 t_{\pi/2}}{\pi t_c}\sum_{i, j} J_{ij}/\hbar\left[
S_{z}^{i}S_{y}^{j}+S_{y}^{i}S_{z}^{j} +S_{x}^{i}S_{y}^{j}+S_{y}^{i}S_{x}^{j}\right].
\label{Eq.Hamiltonian_correction}
\end{equation}

In the framework of Average Hamiltonian Theory, this correction corresponds to the zeroth-order average Hamiltonian acting during the pulses \cite{Choi2020} which in our case results in additional two-body interaction cross terms weighted by the ratio of the pulse duration over the cycling time $t_{\pi/2}/t_c$. 
Figure~\ref{fig:FigSnumerics}B present the three different numerical simulations of the dynamics of $H(t)$, $H_{\text{XXX}}$ and $H_{\text{XXX,corr}}$ in blue, green and yellow respectively. We see that the addition of the correction cross terms of Eq. \ref{Eq.Hamiltonian_correction} is capturing partially the dynamics observed in the simulation of $H(t)$. The remaining differences could be captured by evaluation of the higher order Magnus expansion terms as well as their associated corrections following the derivations of the reference \cite{Choi2020}.

\subsection*{The effect of the four-pulse sequence on a two-particle system}

The effect of the four-pulse sequence on a system interacting under the XX Hamiltonian (Eq. \ref{Eq.H(t)}) and initialized in $\ket{\rightarrow}_x^{\otimes N}$ can intuitively be understood for the simplified\footnote{The following picture does not require any assumption on interaction strength $J$. This is only the case because we consider two spins in this specific example. In addition, we assume delta-like pulses here.} case of two spins ($N=2$), similar to the picture given in reference \cite{Ye2013}. For this case, the initial state after initialization reads
\begin{equation}
\ket{\psi_0} = (\frac{1}{\sqrt{2}} (\ket{\uparrow} + \ket{\downarrow}))\otimes(\frac{1}{\sqrt{2}} (\ket{\uparrow} + \ket{\downarrow})) = \frac{1}{2} (\ket{\uparrow \uparrow} + \ket{\downarrow \downarrow} + \ket{\uparrow \downarrow}+\ket{\downarrow \uparrow}).
\label{Eq.2Spin_state}
\end{equation}
$\ket{\uparrow \uparrow}$ and $\ket{\downarrow \downarrow}$ are both eigenstates of the XX Hamiltonian $H_{\rm{XX}} = J  \left(S_{x}^{1}S_{x}^{2} + S_{y}^{1}S_{y}^{2} \right)$ associated with the same energy eigenvalue $E_1 = 0$. To simplify the following calculations we define the super-position states $\ket{\alpha} = \ket{\uparrow \uparrow} + \ket{\downarrow \downarrow}$ and $\ket{\beta} = \ket{\uparrow \uparrow} - \ket{\downarrow \downarrow}$. Furthermore $\frac{\ket{C}}{\sqrt{2}} = \frac{1}{\sqrt{2}}(\ket{\uparrow \downarrow}+\ket{\downarrow \uparrow})$ constitutes an eigenstate with energy $E_2 = J$ such that the initial state reads $\ket{\psi_0} = \frac{1}{2} (\ket{\alpha}  + \ket{C})$. 

The effect of $\pi/2$ pulses in the $x-$ and $y-$direction can be seen as an operation on the state. The important operation of the pulse in $x-$direction is an effective swap of the two-particle states $\ket{\alpha}$ and $\ket{C}$. The operation performed by the $\pi/2$ pulse in  $y-$direction does not change $\ket{\alpha}$ but transforms $\ket{C}$ to $\ket{\beta}$.

In the following, we present the effect of the whole pulse sequence associated to the accumulated phase acquired during the respective evolution times. 

\begin{itemize}
    \item During $\tau_1$ : $\ket{\psi} = \frac{1}{2} (\ket{\alpha}  + e^{i\frac{J}{\hbar}\tau_1} \ket{C})$
    \item $\pi/2_x$ pulse : $\ket{\psi} = \frac{1}{2} (\ket{C}  + e^{i\frac{J}{\hbar}\tau_1} \ket{\alpha})$
    \item During $\tau_2$ : $\ket{\psi} = \frac{1}{2} (e^{i\frac{J}{\hbar}\tau_2} \ket{C}  + e^{i\frac{J}{\hbar}\tau_1} \ket{\alpha})$
    \item  $\pi/2_{-y}$ pulse : $\ket{\psi} = \frac{1}{2} (e^{i\frac{J}{\hbar}\tau_2} \ket{\beta}  + e^{i\frac{J}{\hbar}\tau_1} \ket{\alpha})$
    \item During $\tau_3$ : $\ket{\psi} = \frac{1}{2} (e^{i\frac{J}{\hbar}\tau_2} \ket{\beta}  + e^{i\frac{J}{\hbar}\tau_1} \ket{\alpha})$ (no phase accumulation*)
    \item  $\pi/2_{y}$ pulse : $\ket{\psi} = \frac{1}{2} (e^{i\frac{J}{\hbar}\tau_2} \ket{C}  + e^{i\frac{J}{\hbar}\tau_1} \ket{\alpha})$
    \item During $\tau_2$ : $\ket{\psi} = \frac{1}{2} (e^{i\frac{2J}{\hbar}\tau_2} \ket{C}  + e^{i\frac{J}{\hbar}\tau_1} \ket{\alpha})$
    \item $\pi/2_{-x}$ pulse :  $\ket{\psi} = \frac{1}{2} (e^{i\frac{2J}{\hbar}\tau_2} \ket{\alpha}  + e^{i\frac{J}{\hbar}\tau_1} \ket{C})$
    \item During $\tau_1$ :  $\ket{\psi} = \frac{1}{2} (e^{i\frac{2J}{\hbar}\tau_2} \ket{\alpha}  + e^{i\frac{2J}{\hbar}\tau_1} \ket{C})$
\end{itemize}

Thus, the pulses are rotating the states and shift the phase factors over the sequence. In the specific case of $\tau_1 = \tau_2 = \tau_3 = \tau$ (corresponding to an XXX Hamiltonian) the states only accumulate a global phase $e^{i\frac{2J}{\hbar}\tau}$ over the whole sequence. 

We directly compare this dynamic with the evolution under $H_{\rm{XXX}} = J  \left(S_{x}^{1}S_{x}^{2} + S_{y}^{1}S_{y}^{2}+ S_{z}^{1}S_{z}^{2} \right)$. In this specific case, the eigenstates $\ket{\uparrow \uparrow}$, $\ket{\downarrow \downarrow}$ and $\frac{1}{\sqrt{2}}(\ket{\uparrow \downarrow}+\ket{\downarrow \uparrow})$ have all the same energy eigenvalue $E = J$ such that the time evolution of the state Eq. \ref{Eq.2Spin_state} reads $\ket{\psi(t)} = \frac{1}{2} (e^{i\frac{J}{\hbar}t} \ket{\uparrow \uparrow} + e^{i\frac{J}{\hbar}t}\ket{\downarrow \downarrow} + e^{i\frac{J}{\hbar}t}(\ket{\uparrow \downarrow}+\ket{\downarrow \uparrow}).$ The state only accumulates a global phase and thus the sequence mimics the evolution of the XXX Hamiltonian.

*We note that the delay time $\tau_3$ has no effect for the specific case of our initial state $\ket{\psi_0}$. This can also intuitively be understood in the Floquet Hamiltonian picture and Eq. \ref{Eq.HAHT}. Only the $J_y$ and $J_z$ components are affected by $\tau_3$ and the resulting Hamiltonian will always be a XZZ-model for $\tau_1 = \tau_2$ for which $\ket{\psi_0}$ constitutes an eigenstate.

\subsection*{Realistic scenario of efficient Hamiltonian Engineering}

Here we propose a setting that would be appropriate to realize highly efficient Hamiltonian Engineering under conditions shown in Fig. 2B, i.e. Rabi frequencies above $\Omega/2\pi = 0.5$ GHz. Technically, such high Rabi frequencies and the timing of $\pi/2$ pulses can be achieved using commercially available microwave technologies. However, for strong driving the two-level approximation for our two Rydberg states might not be fulfilled because the microwave also couples to the surrounding Zeeman states $\ket{48P_{3/2},m_j = -1/2}$ and $\ket{48P_{3/2},m_j = 3/2}$ due to imperfect polarization. We want to show that the coupling and population transfer to the surrounding states can be kept small enough even for Rabi frequencies above 0.5 GHz.

We have determined that 97\% of the microwave power couples to the desired $\ket{48S_{1/2},m_j = 1/2}$ - $\ket{48P_{3/2},m_j = 1/2}$ transition. Thus, a Rabi frequency of $\Omega/2\pi = 0.5$ GHz would lead to a couplings of $\Omega_{\sigma_-}/2\pi = 45$ MHz and  $\Omega_{\sigma_+}/2\pi = 75$ MHz on the $\ket{48S_{1/2},m_j = 1/2}$ - $\ket{48P_{3/2},m_j = -1/2}$ and $\ket{48S_{1/2},m_j = 1/2}$ - $\ket{48P_{3/2},m_j = 3/2}$ $\sigma_-$ and $\sigma_+$ transition respectively. Those transitions are detuned compared to the main one by the presence of the magnetic field resulting in a Zeeman splitting of 1.86 MHz/G. The percentage of the population coupled to these unwanted states is given by $\Omega_{\sigma^{+/-}}^2/(\Omega_{\sigma^{+/-}}^2+\Delta^2)$.
In order the keep this value below $5\%$ we would require a magnetic field of at least B = 333 G.

\end{document}